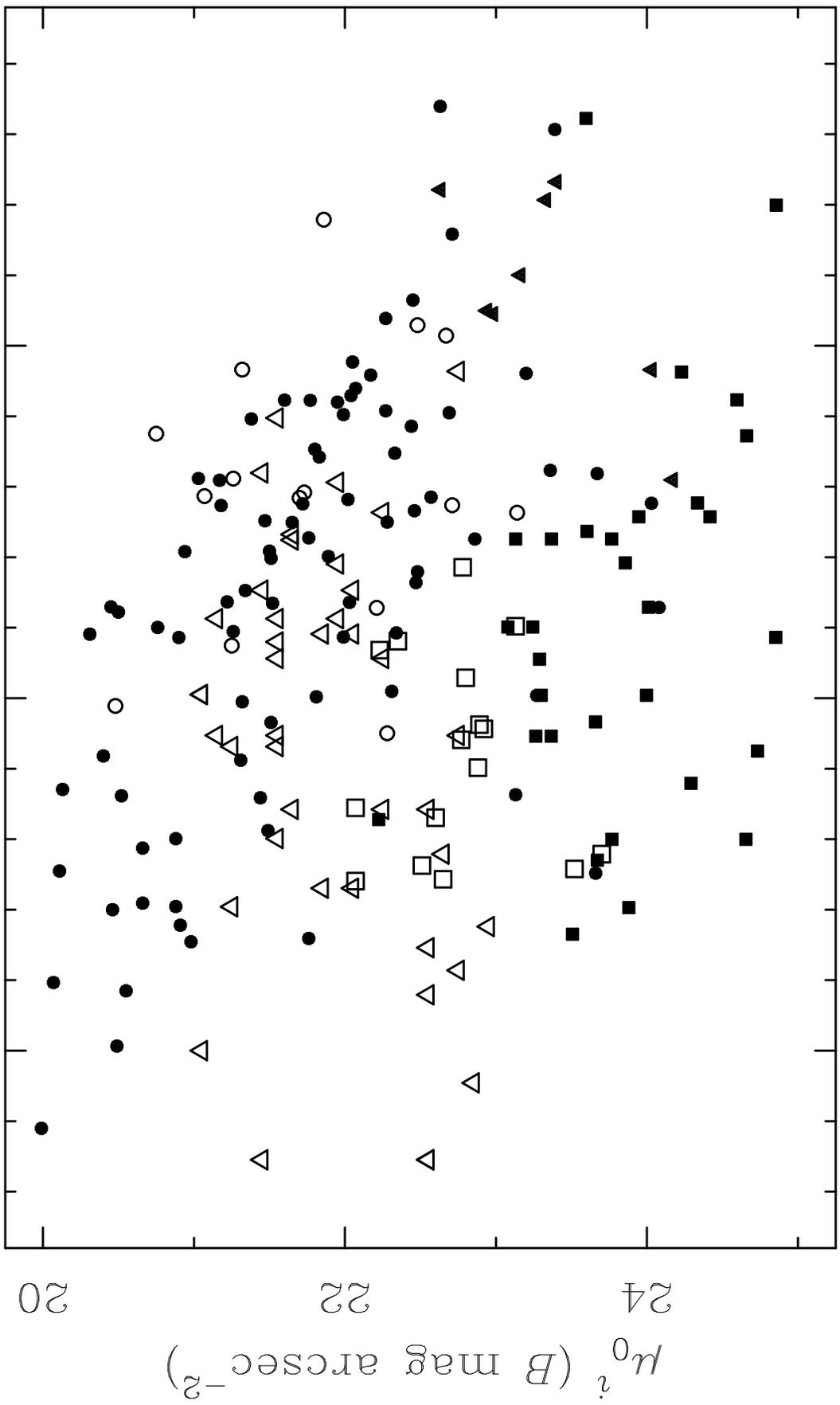

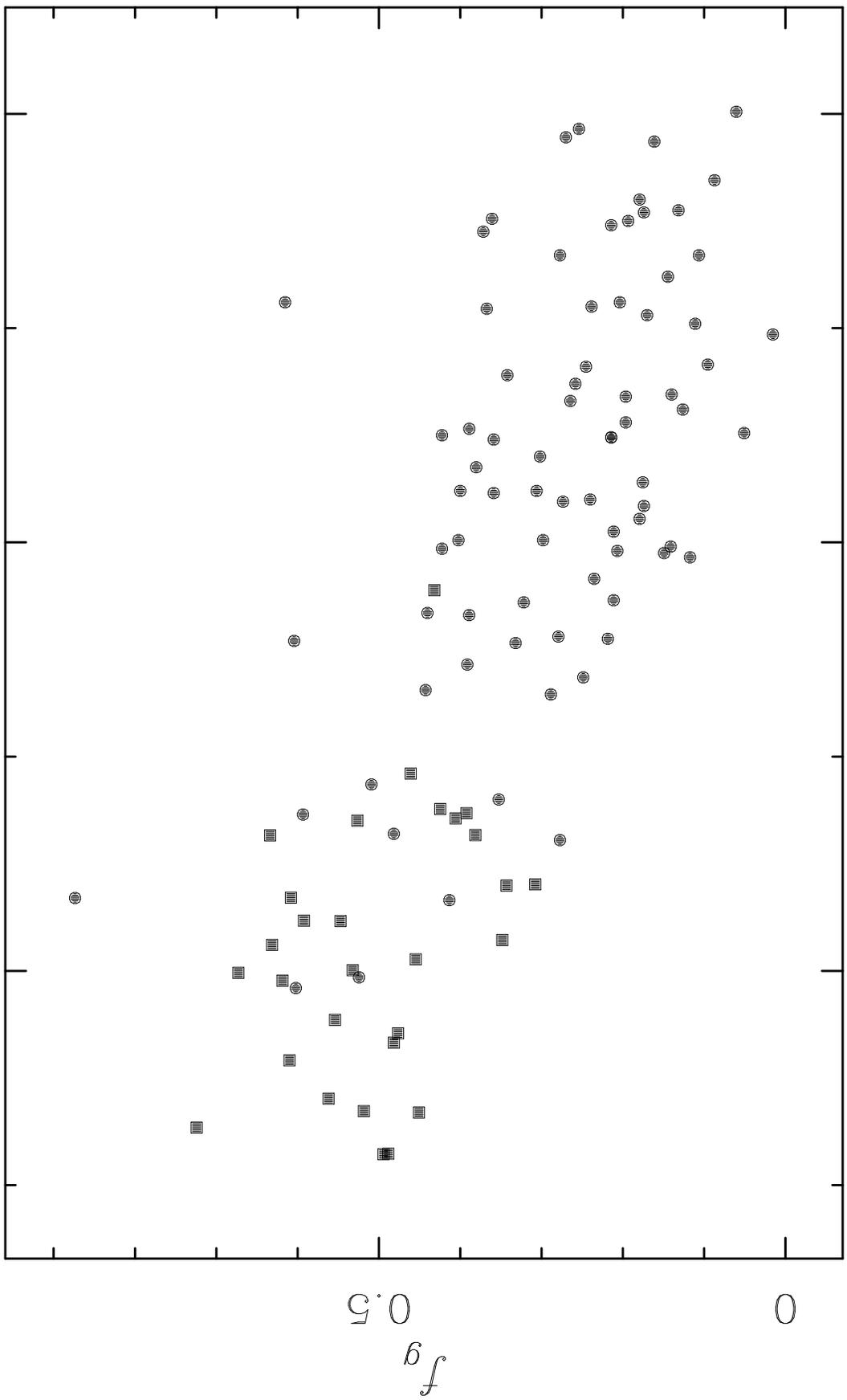

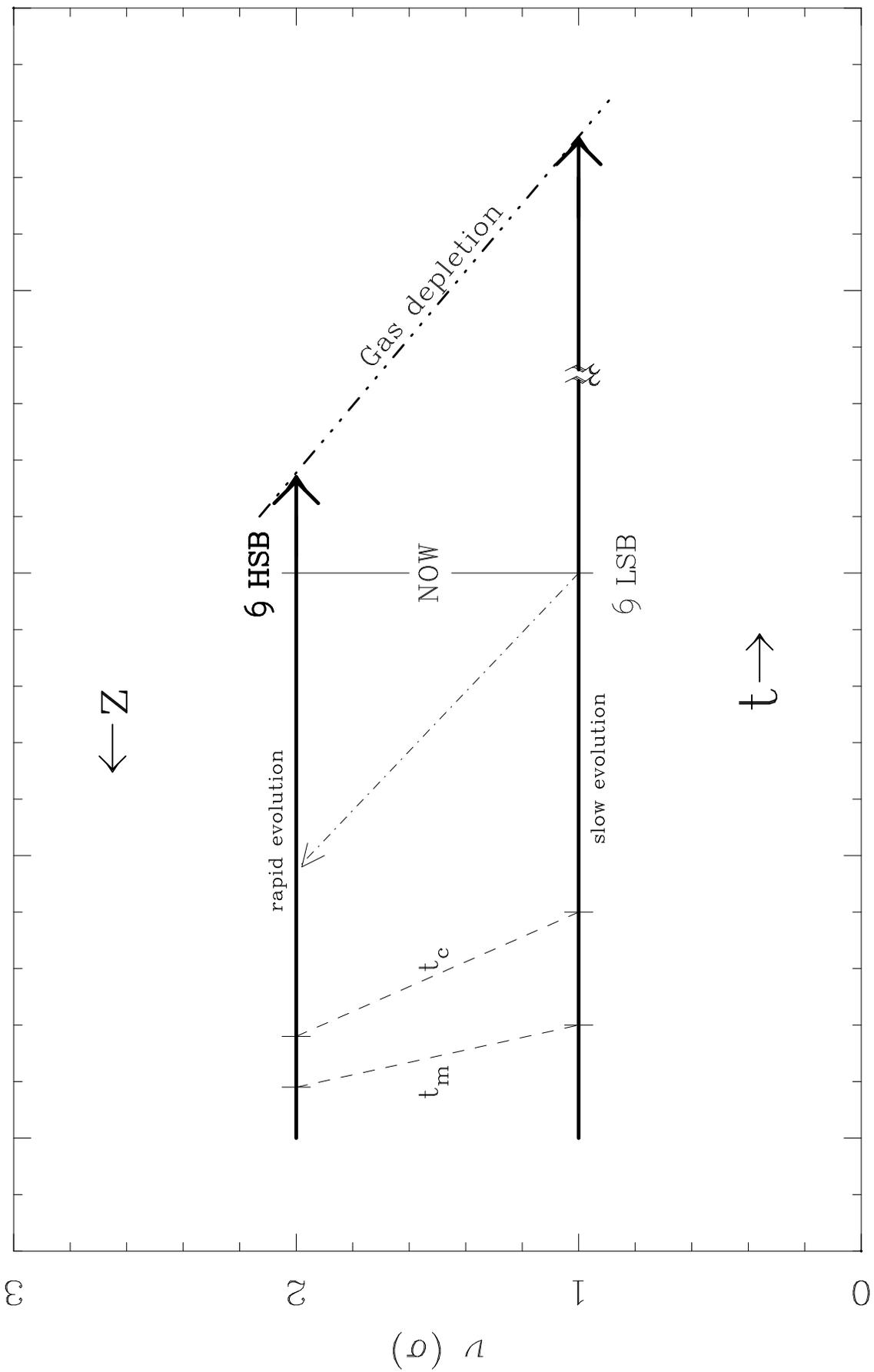

# DWARF AND LOW SURFACE BRIGHTNESS GALAXIES

*Field Populations*

STACY S. MCGAUGH
*Institute of Astronomy*
*The Observatories*
*Cambridge, England*

## 1. Introduction

The terms 'dwarf' and 'low surface brightness' are commonly used to mean a variety of different things, and are sometimes used interchangeably. It is thus necessary to explicitly quantify the definitions which are adopted. Here I shall limit the discussion to field galaxies which are [exponential] disks characterized by a scale size $h$ and central surface brightness $\mu_0$.

The word dwarf implies things which are very small in linear extent, so one definition might be $h < 1$ kpc. More commonly, dwarf is used to mean things which are intrinsically faint, $L < 0.01 L^*$. By either criterion, such things are extremely rare in field samples (Fig. 1), simply because they are so faint that surveys are sensitive to them only over very limited volumes.

On the other hand, there exist many low surface brightness (LSB) disk galaxies which are not small, and do not satisfy either of the above definitions of 'dwarf'. Indeed, these exist right up to $L^*$ (Fig. 1), and typically exhibit spiral structure (McGaugh et al. 1995a; de Blok et al. 1995a). So I use the term LSB to refer to disks with $\mu_0 > 22.5$ $B$ mag arcsec$^{-2}$ (about the brightnesses of the moonless sky) which are too large and luminous to be considered dwarfs. This population constitutes $\sim 1/2$ of all disks by number (McGaugh et al. 1995b; Sprayberry et al. 1995a).

## 2. Dwarf Galaxies

While a great deal is known about dwarfs in the local group (Hodge 1971; 1989) and in nearby groups and clusters (Ferguson & Binggeli 1994), little is known about true dwarfs in the field for the simple reason that their faintness makes them very rare in flux limited samples. I will therefore further limit the discussion to the population of dwarfs discovered by their strong emission lines. Variously known as H II galaxies, BCDs, & BCGs (see Salzer et al. 1989), these are intrinsically small galaxies (though of course with some distribution of sizes) undergoing a strong burst of star formation involving typically $\sim 10^4$ O stars.

These galaxies have been extensively studied both for their remarkable star formation activity, and because they dominate objective prism surveys (Salzer et al. 1989; Terlevich et al. 1991). They are typically low metallicity ($Z \approx 0.2 Z_\odot$), gas rich objects, and as such are relatively unevolved. This led to the hope that some might be true protogalaxies undergoing their first episode of star formation, but with the famous exception of I Zw 18, so far all do have older underlying stellar populations (Salzer, private communication; Telles 1995).

Since this conference focuses on evolution, which primarily means the star formation history (Kennicutt 1995), I would like to review an important outstanding problem these objects pose: the progenitor problem (e.g., Tyson & Scalo 1988). This arises because the inferred star formation history is one of brief episodic bursts interspersing lengthy quiescent periods. Therefore, there must be a vast reservoir of progenitors for each individual H II galaxy currently undergoing a burst:

$$n_{prog} = \frac{\tau_{off}}{\tau_{on}} n_{burst}$$

where $n_{burst}$ is the observed density of currently active galaxies, and $\tau$ represents the duty cycle for star formation, i.e., the period of time spent in bursting and quiescent phases. The quiescent phase is essentially a Hubble time less the bursts, which are generally inferred to be few and brief, $\tau_{on} \sim 10^7$ yr. Thus

$$n_{prog} \approx \frac{10^{10}}{10^7} n_{burst} \approx 1000 \, n_{burst}$$

which is an enormous problem since $n_{burst}$ is observed to constitute $\sim 10\%$ of the total field galaxy population (Salzer 1989; see also Schade & Ferguson 1994). Thus, for any amount of fading after the burst, the inferred number of progenitors exceeds everything else we know about optically by a factor of $\sim 100$, but is undetected in 21 cm surveys (Weinberg et al. 1991).

One way to ease this problem is to increase $\tau_{on}$, presumably with a concomitant decrease in burst strength consistent with recent estimates that the burst itself contributes $\lesssim 1$ mag. to the total luminosity (Salzer, private communication). This then leads to a qualitatively different picture for the star formation history with substantial peaks and troughs but not sharp $\delta$-functions. However, one is limited in the degree to which the star formation rate can be smoothed out in this sense by the need to

1. not overproduce luminosity in long lived stars
2. not overproduce metallicity
3. not consume all the gas, and most crucially
4. provide enough ionizing photons to yield the observed H$\alpha$ luminosities for long periods.

Item [1] could be avoided by truncating the IMF so that only high mass ($\gtrsim 10\,M_\odot$) stars are formed in the burst. While this is appealing in some respects, tailor made IMFs can fit anything and there really is no evidence for variations in the IMF (McGaugh 1991). Metallicity [2] may be lost in preferentially enriched supernova driven winds, but note that there is no evidence that these galaxies 'explode' and lose *all* of their gas. Item [3] provides the ultimate constraint unless very substantial amounts of gas are subsequently accreted to replenish the supply. I think item [4] places the tightest constraints on the burst duration, but given the desperate lack of adequate model atmospheres for hot, low metallicity stars, it is conceivable that not quite so many O stars are required if low metallicity stars produce a lot more ionizing photons per unit mass than is usually assumed based on solar metallicity models.

Another, related puzzle is that the *underlying* stellar population is itself very blue (Telles 1995). For a star formation history consisting of a few intermittent bursts, the remnants of the preceding burst should have reddened substantially. This appears not to be the case, and the colors are so blue ($B - V \approx 0.4$) that I don't think that low metallicity can be the entire explanation. It is also hard to see how to address this by varying the IMF, since we are considering an underlying population which is presumably much older than the lifetimes of blue stars. Perhaps the mean age is implicated — either it has not been long since the previous burst, or the system as a whole formed late and is rather less than 10 Gyr old, or most likely some combination of these and metallicity effects.

A young mean age for the underlying population suggests a decrease in $\tau_{off}$, but even if one takes $\tau_{on} \to$ a few $\times\,10^8$ and $\tau_{off} \to$ a few $\times\,10^9$, the entirety of the normal galaxy population fainter than $\lesssim \frac{1}{2}L^*$ is needed to serve as progenitors. A substantive progenitor population seems to be demanded by the intensity of the observed star formation, but ruled out by optical and 21 cm surveys. However, optical surveys are very insensitive to objects which are both small and low surface brightness, and these could be quite *numerous* without violating the 21 cm constraints on *mass* density.

## 3. Low Surface Brightness Galaxies

Now let us turn to the population of LSB galaxies which are comparable in size to the high surface brightness spirals which define the Hubble sequence. These galaxies are extremely blue ($B - V \approx 0.4$, $V - I \approx 0.7$), especially in the redder colors (McGaugh & Bothun 1994; Rönnback & Bergvall 1994; de Blok et al. 1995a). The colors of disks become generally bluer (with much scatter) as either size or surface brightness decrease. This suggests a connection between small, LSB galaxies and the underlying components of H II galaxies, but in this section I will discuss larger, Milky Way size objects which are unlikely to contribute to the progenitor population.

Understanding the very blue colors of LSB galaxies is challenging. In order to disentangle the effects of age and metallicity, it is useful to measure the latter. LSB galaxies are quite metal poor, with typical metallicities in the range $0.1 < Z < 0.3 Z_\odot$ (McGaugh 1994; Rönnback & Bergvall 1995). Thus at least some of the blueness must be due to this. However, metallicity can not explain it entirely, as color and metallicity are not correlated (McGaugh & Bothun 1994). A rather low mean age is thus implicated, with a birth rate function weighted more heavily towards recent epochs than early ones. This is at once consistent and at odds with the trends along the Hubble sequence (Kennicutt 1995): as morphological types typically later than Sc, one might expect LSB galaxies to have such birth rate functions. However, they also have low current star formation rates per unit area, as in very early type disks.

The inferred ages are typically a few Gyr less than those of high surface brightness disks, suggesting a late formation epoch and/or slow evolution. The latter is certainly indicated by the low metallicities, and also by the large gas mass fractions (Fig. 2). Given the observed ratio of 21 cm to optical flux ($M_{HI}/L$), the gas mass fraction $f_g = M_{gas}/(M_{gas} + M_*)$ can be calculated from

$$f_g = \left[1 + \frac{\Upsilon_*}{\eta} \frac{L}{M_{HI}}\right]^{-1}$$

with some reasonable assumption about the stellar mass to light ratio $\Upsilon_*$ and the fraction of gas in H I, $\eta^{-1}$. For simplicity, I take $\Upsilon_* = \eta$; though not exactly correct since LSB galaxies have little molecular gas (Schombert et al. 1990), this is not a bad approximation and the trend is clear in the raw data (de Blok et al. 1995b; Sprayberry et al. 1995c).

The correlation between $f_g$ and $\mu_0$ is remarkably strong, but there is no correlation between $f_g$ and scale length. The lack of objects with high gas fractions at high surface brightnesses is certainly real and can not be a selection effect. The lack of galaxies with low $f_g$ at low surface brightnesses could very well be a selection effect, in which case the correlation line would be the upper envelope of the distribution. While there certainly exist dwarf Elliptical galaxies in this regime, it is important to determine if these are causally connected populations. That is, do spirals evolve to lower $f_g$ at fixed $\mu_0$, perhaps also evolving in morphology, or along the sequence to higher $\mu_0$? I suspect the latter, but of course some combination is possible.

The trend of global gas fraction seen in Fig. 2 is also mimiced locally: low surface brightness stellar disks have low surface density H I disks. However, the H I density does not vary over as large a range as that in optical surface brightness: for a factor of five change in surface brightness, the H I surface density changes by a factor of only $\sim 2$. This holds the key to the inhibited evolutionary rates of LSB galaxies: they exist close to the critical threshold for star formation (Kennicutt 1989), and as a result form stars at a very slow rate in spite of their enormous gas reservoirs. (Note that the usual assumption that the star formation rate is proportional to the gas mass, which leads to exponential star formation histories with $\dot{M} \propto e^{-t/\tau_s}$ can not hold in these galaxies unless $\tau_s \leq 0$, i.e., an *increasing* star formation rate or a constant one with a low age.)

Surface density is thus a critical parameter in governing a disk's evolution. So what determines the density? All lines of evidence, the low metallicities, blue colors, and large gas mass fractions, indicate slow evolution and relative youth. One expects a galaxy to form late if it arises from a low density peak in the initial field of fluctuations. That low initial density should lead directly to a low final density (Fig. 3), with the observed consequences. This simple picture, derived from the physical properties listed above (McGaugh 1992), makes a clear prediction about the spatial distribution of LSB galaxies: they should be less strongly clustered than higher surface brightness spirals. This prediction has been confirmed (Mo et al. 1994) with the additional observation that LSB galaxies are extremely isolated on small ($< 2$ Mpc) scales (see also Bothun et al. 1993). They have no bright companions, and have suffered no tidal perturbations which might clump their gas and induce star formation (presumably raising their surface brightnesses). They have endured no merging, being the poster children for galaxy formation by gradual collapse. They may, however, compose a population which, in hierarchical structure formation scenarios, is expected to fall into larger group and cluster structures at late times (Rakos & Schombert 1995).

The evolution of disks is governed by their characteristic density as well as total mass. The surface brightness of a disk is intimately related to its evolutionary rate and collapse epoch. The star formation history is relatively stable in large disks, but tends increasingly towards episodic bursts as size decreases. This may simply be a statement that star formation is inherently a local process, so that larger disks in effect average over larger numbers of discrete star forming events.

**DISCUSSION:**

**McCall**: In your graphs of metallicity vs. color; were the colors integrated, or were they corrected for the star forming regions? I am concerned that the scatter in the colors might be in part due to the effect of the young component, which of course can vary substantially in luminosity fraction from galaxy to galaxy.

**McGaugh**: For H II galaxies a correction must very carefully be applied; it is not necessary for LSB galaxies.

**Meurer**: How do you know that the cycle time in BCDs is $10^7$ years? From the size of the "starburst" region I would say this is more like a lower limit to the cycle time, although the clusters they contain should themselves be better analogs to true instantaneous bursts.

**McGaugh**: A burst duration of $\sim 10^7$ yr is the consensus number in the literature, though I have some sympathy for the case that it be longer.

**Djorgovski**: These objects obviously have fewer stars than high surface brightness galaxies, but for a given type (e.g., giant disks, true dwarfs, etc.), do they have fewer baryons, and do they have less dynamical mass?

**McGaugh**: Fewer stars per unit area, certainly. They do not have much lower dynamical masses; the mass to light ratio within the optical radius increases very systematically with decreasing surface brightness (Zwaan et al. 1995; de Blok et al. 1995b). Whether this is due to fewer baryons per unit mass or what is very hard to say, and poses a fundamental puzzle.

*Figure 1.* Central surface brightness vs. scale length for disk galaxies [data from Boroson (1981; open circles), Romanishin et al. (1983; open squares), van der Kruit (1987; open triangles), McGaugh & Bothun (1994; solid squares), de Blok et al. (1995a; solid squares), Sprayberry et al. (1995b; solid triangles), and de Jong (1995; solid circles)]. Galaxies fill this plane up to maxima in both luminosity and surface brightness.

*Figure 2.* The correlation of gas mass fraction with surface brightness.

*Figure 3.* The surface brightness of a disk is related to the amplitude $\nu$ of the primordial density perturbation from which it arises. Low density perturbations collapse late ($t_c$), forming low density galaxies which evolve slowly. When observed at any given epoch, their gas content, metallicity, etc. will correspond to an earlier state of higher density galaxies (dash-dotted line).